\documentclass[a4paper,11pt]{article}

\usepackage{pos}
\usepackage{braket}

\usepackage[T1]{fontenc} 
\usepackage{multirow}
\usepackage{tikz}

\usepackage{pdflscape}
\usepackage{booktabs}
\usepackage{verbatim}
\usepackage{amsmath}
\usepackage{graphics}
\usepackage{mathtools}
\usepackage{slashed}
\usepackage{subcaption}

\def\eq#1{Eq.~(\ref{#1})}

\newcommand{\secn}[1]{Section~\ref{#1}}

\newcommand{\be}{\begin{equation}}
\newcommand{\ee}{\end{equation}}
\newcommand{\bea}{\begin{eqnarray}}
\newcommand{\eea}{\end{eqnarray}}
\newcommand{\nn}{\nonumber}

\newcommand{\as}{\alpha_s}
\newcommand{\eps}{\epsilon}
\newcommand{\sig}{\sigma}

\newcommand{\beq}{\begin{eqnarray}}
\newcommand{\eeq}{\end{eqnarray}}
\newcommand{\npo}{{n+1}}
\newcommand{\npt}{{n+2}}
\newcommand{\npth}{{n+3}}
\newcommand{\pn}{\Phi_n}
\newcommand{\pnpo}{\Phi_\npo}
\newcommand{\pnpt}{\Phi_\npt}

\newcommand{\NLO}{{\mbox{\tiny{NLO}}}}
\newcommand{\NNLO}{{\mbox{\tiny{NNLO}}}}
\newcommand{\NNNLO}{{\mbox{\tiny{N$^3$LO}}}}

\newcommand{\CG}{{\mbox{\tiny{CG}}}}

\newcommand{\one}{\, (\mathbf{1})}
\newcommand{\two}{\, (\mathbf{2})}
\newcommand{\three}{\, (\mathbf{3})}
\newcommand{\otwo}{\, (\mathbf{12})}

\newcommand{\RV}{\, (\mathbf{RV})}

\newcommand{\RRV}{\mathbf{RRV}}
\newcommand{\RVV}{\mathbf{RVV}}

\def\eq#1{Eq.~(\ref{#1})}
\def\bra#1{%
  \left\langle\smash{#1}{\vphantom1}\right|}
\def\ket#1{%
  \left|\smash{#1}{\vphantom1}\right\rangle}

\newcommand{\bS}[1]{{\bf S}_{#1}}

\title{Strongly-ordered infrared limits for subtraction counterterms from factorisation}
\ShortTitle{Strongly-ordered counterterms}

\author*[a]{Lorenzo Magnea}
\author[a]{Calum Milloy}
\author[b]{Chiara Signorile-Signorile}
\author[a]{Paolo Torrielli}
\author[a]{Sandro Uccirati}

\affiliation[a]{Dipartimento di Fisica, Universit\`a di Torino, and INFN, Sezione di Torino\\
  Via Pietro Giuria, 1 I-10125, Torino, Italy}

\affiliation[b]{Institut f{\"u}r Theoretische Teilchenphysik, Karlsruher Institut f{\"u}r 
Technologie (KIT), D-76128 Karlsruhe, Germany, and Institut f{\"u}r Astroteilchenphysik, 
Karlsruher Institut f{\"u}r Technologie (KIT), D-76021 Karlsruhe, Germany}

\emailAdd{lorenzo.magnea@unito.it}
\emailAdd{calumwilliam.milloy@unito.it}
\emailAdd{chiara.signorile-signorile@kit.edu}
\emailAdd{paolo.torrielli@unito.it}
\emailAdd{sandro.uccirati@unito.it}

\abstract{After a brief introduction to the problem of subtraction of infrared divergences
for high-order collider observables, we present a preliminary study of strongly-ordered
soft and collinear multiple radiation from the point of view of factorisation. We show that 
the matrix elements of fields and Wilson lines that describe soft and collinear radiation
in factorised scattering amplitudes can be re-factorised in strongly-ordered limits, 
providing a systematic method to compute them, to characterise their singularity 
structure, and to build local subtraction counterterms for strongly-ordered configurations.
Our results provide tools for a detailed organisation of subtraction algorithms, in principle  
to all orders in perturbation theory.}

\FullConference{%
  Loops and Legs in Quantum Field Theory - LL2022,\\
  25-30 April, 2022\\
  Ettal, Germany
}

\def\@preprint{\@empty}
\newcommand\preprint[1]{\gdef\@preprint{\hfill #1}}
\preprint{TTP22-059, P3H-22-096}

\begin{document}
\noindent\@preprint\par
\maketitle


\section{Setting the stage}
\label{Intro}

High-order calculations of collider observables are of paramount importance
in order to achieve the precision goals required by present and forthcoming 
experimental results, and possibly identify new physics signals. In this context,
efficient and general algorithms to implement the cancellation of infrared
divergences are a crucial tool. The problem first arises at next-to-leading 
order (NLO), and in that context efficient and completely general subtraction 
algorithms were developed in the past decades~\cite{Frixione:1995ms,
Catani:1996vz,Nagy:2003qn,Bevilacqua:2013iha}, and are currently 
implemented in the simulation codes employed by LHC experiments. 
The extension of these algorithms to next-to-next-to-leading order (NNLO)
and beyond has been a major research goal for a number of years:
several different methods have been proposed, and many have been
successfully implemented to derive predictions for many LHC processes 
(methods are described, for example, in~\cite{GehrmannDeRidder:2005cm,
Somogyi:2005xz,Czakon:2010td,Binoth:2000ps,Anastasiou:2003gr,
Caola:2017dug,Catani:2007vq,Boughezal:2015dva,Cacciari:2015jma,
Sborlini:2016hat,Herzog:2018ily,Magnea:2018hab,Capatti:2019ypt},
and were recently reviewed in~\cite{TorresBobadilla:2020ekr}).

All subtraction methods rely upon the infrared factorisation properties of 
scattering amplitudes, as applying to both virtual corrections and to soft
and collinear real radiation (for a recent review, see~\cite{Agarwal:2021ais}).
It is natural to imagine that a detailed understanding of factorisation should
play a crucial role in the ultimate organisation of subtraction algorithms, in
principle to any order in perturbation theory. This viewpoint was pursued
in~\cite{Magnea:2018ebr} (see also~\cite{Feige:2014wja}), where general
expressions for soft and collinear counterterms were proposed, valid to
all orders in perturbation theory. 

A significant source of complexity in the organisation of infrared counterterms
beyond NLO is the necessity to properly identify and account for strongly-ordered
soft and collinear configurations, where clusters of particles become unresolved
in a hierarchical pattern. While these configurations are in principle `simple',
and expected to correspond to products of lower-order radiative counterterms,
they must be precisely controlled, since they serve to cancel singularities
arising in mixed real-virtual contributions. Furthermore, the number of 
strongly-ordered configurations (and of the related counterterms) grows
rapidly with the perturbative order.

In the present contribution, we will begin a discussion of strongly-ordered
subtraction counterterms from the point of view of infrared factorisation. We 
will show in a few examples that the matrix elements of fields and Wilson lines
that describe the factorised emission of soft and collinear particles can be
re-factorised in strongly-ordered configurations. This provides formal expressions
for strongly-ordered counterterms to all orders in perturbation theory, and
allows to identify the patterns of cancellation between these counterterms 
and singularities involving mixed real and virtual corrections.

To set the stage, we begin by briefly reviewing the structure of infrared 
subtraction at NLO, in the case of final-state massless particles, and then 
we will show how the problem of strongly-ordered configurations arises at 
NNLO and beyond. While much of the discussion below is general, in 
concrete cases we will use the framework of Local Analytic Sector 
subtraction, developed and discussed in~\cite{Magnea:2018hab,
Magnea:2018ebr,Magnea:2020trj,UcciratiLL}.

At NLO, the subtraction problem can be succinctly summarised. For any 
infrared-safe observable $X$, the NLO distribution is computed by
combining virtual and real corrections, as
\beq
  \frac{d \sigma_\NLO}{d X} \, = \, \lim_{d \to 4} \, \bigg\{ 
  \int d \Phi_n \, V_n \, \delta_n (X) + \int d \Phi_\npo \, R_\npo \, \delta_\npo (X) 
  \bigg\} \, ,
\label{NLOstart}
\eeq
where we assumed that the Born contribution to the observable $X$ 
involves $n$ particles, $V_n$ is the virtual correction to the $n$-particle 
process, $R_\npo$ is the squared matrix element for single real radiation, 
and $\delta_p (X) = \delta (X - X_p)$ fixes the expression for the observable 
in the $p$-particle phase space to the prescribed value $X$. By general 
theorems (see~\cite{Agarwal:2021ais}), explicit infrared poles in $V_n$ are 
cancelled by the integration of phase space singularities for unresolved 
radiation in the second term of \eq{NLOstart}. In order to allow for an
efficient numerical integration of $R_\npo$, the subtraction approach 
proposes to define a {\it local counterterm} $K^{\one}_\npo$, and subsequently 
integrate it over the unresolved particle phase space, according to
\beq
  K^{\one}_\npo \, = \, {\bf L}^{\one} \, R_\npo \, , \qquad \quad 
  I^{\one}_n = \int d \Phi^\npo_{{\rm r}, \, 1} \, K^{\one}_\npo \, ,
\label{NLOcount}
\eeq
where the operator ${\bf L}^{\one}$ extracts the contributions of all singular 
regions to $R_\npo$, and one defines the radiative phase space by $d \Phi_\npo 
= d \Phi_n \, d \Phi^\npo_{{\rm r}, \, 1}$. One may then rewrite \eq{NLOstart} 
identically as
\beq
  \frac{d \sigma_\NLO}{d X} \, = \,
  \int d \Phi_n \Big( V_n + I^{\one}_n \Big) \, \delta_n (X) + 
  \int d \Phi_\npo \Big( R_\npo \, \delta_\npo (X) -  K^{\one}_\npo \delta_n (X) \Big) \, ,
\label{NLOfin}
\eeq
where use has been made of the IR safety of $X$. Now all IR poles cancel 
in the first integral, while phase-space singularities cancel in the second integral,
which can then be performed directly in $d=4$. Clearly, any actual implementation
of this idea relies upon precise definitions of the {\it subtraction operator} 
${\bf L}^{\one}$, and of the {\it phase-space measure} $d \Phi^\npo_{{\rm r}, \, 1}$:
indeed, the $(\npo)$-particle phase space $d \Phi_\npo$ does not naturally 
factorise as suggested, except in the soft limit. Subtraction methods differ
in how these two problems are approached. In the context our method, the 
NLO subtraction operator is defined by partitioning the phase space into {\it sectors}, 
along the lines of~\cite{Frixione:1995ms}, and a proper factorisation of the 
radiative phase space is achieved by employing {\it phase space mappings},
following~\cite{Catani:1996vz} (for a detailed description, see~\cite{Magnea:2018hab}).
A careful choice of different mappings for different contributions to the NLO
local counterterm allows for a straightforward analytic integration, as discussed 
in~\cite{Magnea:2020trj}.

At NNLO, the complexity of the subtraction problem grows very steeply in several 
directions. Phase-space mappings at NNLO and beyond are discussed in detail for
example in Ref.~\cite{DelDuca:2019ctm}, while the extension of our approach to 
NNLO was presented in~\cite{Magnea:2018hab,Magnea:2020trj,UcciratiLL}. 
Here, we will only discuss the general structure of counterterms, in order to highlight 
the appearance and relevance of strongly-ordered configurations. At NNLO, 
infrared cancellations require the combination of three terms: double virtual (VV), 
real-virtual (RV) and double real (RR). In analogy with \eq{NLOstart}, one writes
\beq
  \frac{d \sig_\NNLO}{d X} \, = \, \lim_{d \to 4} 
  \Bigg\{ \! \int d \pn \, VV_n \, \delta_n (X) + \int d \pnpo \, 
  RV_\npo \, \delta_\npo (X) 
  + \int d \pnpt \, RR_\npt \, \delta_\npt (X) 
  \Bigg\} .
\label{NNLOstart}
\eeq 
There are {\it four} kinds of singular phase-space configurations in \eq{NNLOstart},
requiring four different local counterterms. We define
\beq
  K^{\one}_\npt = {\bf L}^{\one} \, RR_\npt \, , \, \, \, \,  
  K^{\two}_\npt = {\bf L}^{\two} \, RR_\npt \, , \, \, \, \,  
  K^{\otwo}_\npt = {\bf L}^{\one} \, {\bf L}^{\two}  \, RR_\npt \, , \, \, \, \,  
  K^{\RV}_\npo = \widetilde{\bf L}^{\one} \, RV_\npo \, ,
\label{NNLOcount}
\eeq
where $K^{\one}_\npt$ collects configurations where a single particle becomes 
unresolved, $K^{\two}_\npt$ those where two particles become unresolved, and
$K^{\otwo}_\npt$ defines their overlap: notice that the definition of $K^{\otwo}_\npt$
implies that this counterterm collects precisely the strongly-ordered configurations 
that we are discussing. Finally, $K^{\RV}_\npo$ collects configurations where the 
single radiated particle in the $RV$ contribution becomes unresolved. The 
operator $\widetilde{\bf L}^{\one}$ differs from ${\bf L}^{\one}$ by the inclusion
of a set of subleading-power contributions in the relevant normal variables, whose
role will become apparent below (for details, see~\cite{UcciratiLL}). In analogy with
\eq{NLOcount}, one must then integrate the local counterterms over their radiative 
phase spaces, defining
\beq
  && I^{\one}_\npo = \int d \Phi^\npt_{{\rm r}, \, 1} \, K^{\one}_\npt \, , \quad
  I^{\otwo}_\npo = \int d \Phi^\npt_{{\rm r}, \, 1} \, K^{\otwo}_\npt \, , \quad
  I^{\two}_n = \int d \Phi^\npt_{{\rm r}, \, 2} \, K^{\two}_\npt \, , \nn \\
  && \hspace{4cm} I^{\RV}_n = \int d \Phi^\npo_{{\rm r}, \, 1} \, K^{\RV}_\npo \, .
\label{NNLOintcount}
\eeq
One finally constructs a subtracted expression for the NNLO distribution,
\beq
\label{NNLOfin} 
  \frac{d \sig_\NNLO}{dX} & = & \hspace{-1mm} \int d \Phi_n
  \Big[ VV_n + I^{\two}_n + I^{\RV}_n \Big] \, \delta_n (X) \\
  && \hspace{-2mm} + \, \int d \Phi_\npo \, \bigg[
  \Big( RV_{\npo} + I^{\one}_\npo \Big) \, \delta_\npo (X) \, -  
  \Big( K^{\RV}_\npo + I^{\otwo}_\npo \Big) \, \delta_n (X) \bigg] \nn \\
  && \hspace{-2mm} + \, \int d \Phi_\npt \, \bigg[
  RR_\npt \, \delta_\npt (X) - K^{\one}_\npt \, \delta_\npo (X) - 
  \Big( K^{\two}_\npt - K^{\otwo}_\npt \Big) \delta_n (X) \bigg] \, . \nn
\eeq
In \eq{NNLOfin}, the last line can be integrated directly in $d = 4$, as
all singular configurations have been subtracted from $RR_\npt$ without 
double counting. In the second line, the first parenthesis is free of IR poles
by standard arguments; also, the phase-space singularities of 
$RV_\npo$ are subtracted, by construction, by $K^{\RV}_\npo$, and 
those of $I^{\one}_\npo$ by $I^{\otwo}_\npo$; finally, the inclusion of 
non-minimal terms (which are non-singular in the radiative phase space) 
in $\widetilde{\bf L}^{\one}$  allows to cancel the IR poles of $I^{\otwo}_\npo$. 
When this delicate balance of cancellations is achieved in the second line, 
the first line is bound to be finite by the general theorems enforcing the 
cancellation of IR singularities. We wish to emphasise that the importance
of strongly-ordered configurations grows steeply with the perturbative order:
to this end, we write down the formal extension of \eq{NNLOfin} to N$^3$LO.
Using notations that hopefully generalise transparently the NNLO case,
one can write the subtracted distribution as 
\beq
\label{N3LOfin}  
  \frac{d \sig_\NNNLO}{dX} & = & \int d \Phi_n \, 
  \Big[ VVV_n + I_n^{\three} + I_n^{(\RVV)}  + I_n^{(\RRV, \,{\mathbf 2})} 
  \Big] \, \delta_n (X) \nn \\
  && \hspace{-5mm} + \, \int d \Phi_\npo \,
  \Big( RVV_\npo + I^{\two}_\npo + I^{(\RRV, \, \mathbf{1})}_\npo
  \Big) \, \delta_\npo (X) \, - \,
  \Big( K^{(\RVV)}_\npo + I^{(\mathbf{2 3})}_\npo
  + I^{(\RRV, \, \mathbf{12})}_\npo \Big) \, \delta_n (X) \nn \\
  && \hspace{-5mm} + \, \int d \Phi_\npt \, \bigg\{
  \Big( RRV_\npt + I^{\one}_\npt \Big) \, \delta_\npt (X) -
  \Big( K^{(\RRV, \mathbf{1})}_\npt + I^{\otwo}_\npt \Big) \, \delta_\npo (X)
  \nn \\
  & & \hspace{3cm} - \, \bigg[
  \Big( K^{(\RRV, \, \mathbf{2})}_\npt + I^{(\mathbf{13})}_\npt \Big) -
  \Big( K^{(\RRV, \mathbf{12})}_\npt  + I^{(\mathbf{123})}_\npt \Big)
  \bigg] \, \delta_n (X) \bigg\} \nn \\
  && \hspace{-5mm} + \, \int d\Phi_\npth \bigg[ RRR_\npth \, 
  \delta_\npth (X) - K^{\one}_\npth \, \delta_\npt (X) -
  \Big( K^{\two}_\npth - K^{\otwo}_\npth \Big) \, \delta_\npo (X) \nn \\
  && \hspace{3cm} - \,
  \Big( K^{(\mathbf{3})}_\npth - K^{(\mathbf{13})}_\npth
  - K^{(\mathbf{23})}_\npth + K^{(\mathbf{1 2 3})}_\npth \Big) \,
  \delta_n (X) \bigg] \, .
\eeq
One sees that in order to achieve the cancellation of IR poles and the finiteness
of the radiative phase space integrations in each line of \eq{N3LOfin} one needs
a total of {\it eleven} local counterterms: of these, {\it five} display various kinds 
of strong ordering; for example, $K^{(\mathbf{23})}_\npth$ contains configurations 
with three unresolved particles, out of which two become unresolved faster than 
the third one. Clearly, a systematic analysis of strongly-ordered IR configurations
is warranted, if one wishes to understand the subtraction process to all orders.

\section{Infrared counterterms from factorisation}
\label{Facto}

In Ref.~\cite{Magnea:2018ebr}, a systematic proposal was presented for the
construction of soft and collinear local counterterms to all orders in perturbation
theory. We display here some sample definitions, in order to introduce the
appropriate language for analysing the strongly-ordered case.  For multiple
{\it soft} emissions, one can define {\it eikonal form factors}
\beq
  {\cal S}_{n, \, m} \Big(\{\beta\}, \{k\}, \{\lambda\} \Big) \, = \,  
  \bra{k_1, \lambda_1; \ldots ; k_m, \lambda_m} \, T \bigg[ 
  \prod_{i = 1}^n \Phi_{\beta_i} (\infty, 0) \bigg] \ket{0} \, ,
\label{softrad}
\eeq
where $\Phi_{\beta}(\infty, 0)$ is a semi-infinite Wilson line stretching along 
the direction of the four-vector $\beta$, and $T$ denotes time-ordering. Eikonal 
form factors mimic the emission of a system of $m$ soft particles with momenta 
$k_l$ from $n$ hard emitters with four-velocities $\beta_i$, at leading power 
in the soft momenta. Squaring the form factors, one gets {\it eikonal transition 
probabilities}
\beq
\label{SoftFunc}
  S_{n, m} ( \{ \beta \}, \{ k \} ) & = & \sum_{\lambda_i}
  \braket{0 | \, \overline{T} \bigg[ \prod_{i = 1}^n \Phi_{\beta_i}(0,\infty)\bigg]  
  | \{ k \}, \{ \lambda \} }
   \braket{ \{ k \}, \{ \lambda \} | T \bigg[ \prod_{i = 1}^n \Phi_{\beta_i} (\infty,0) 
   \bigg] | 0 } \, ,
\eeq
which are natural candidates for soft local subtraction counterterms, since 
they retain all leading-power information on soft radiation. Note that the soft 
limit here is taken uniformly for all soft particles. The suitability of \eq{SoftFunc}
as a soft counterterm is supported by integrating over the soft space and 
summing over the number of soft particles. Using {\it completeness}, this 
yields
\beq
  \sum_{m = 0}^\infty \int d \Phi_m \, S_{n, \,m} 
  \Big(\{ \beta \}, \{ k \} \Big) \, = \, 
  \bra{0} \,  \overline{T} \left[ \prod_{i = 1}^n 
  \Phi_{\beta_i} (0, \infty) \right] T \left[ \prod_{i = 1}^n 
  \Phi_{\beta_i} (\infty, 0) \right] \ket{0} \, .
\label{complete}
\eeq
which is a total cross section in the presence of Wilson-line sources, and
thus IR finite by general cancellation theorems. Note that \eq{complete} requires
a uniform treatment of the UV region for different terms, since the phase-space
integration is unbound; in the IR region, on the other hand, $d \Phi_m$ can 
be taken as the product of single-particle phase spaces, since the full phase 
space naturally factorises in the soft limit. A similar analysis can be performed 
for {\it collinear} radiation. In this case one can define {\it radiative jet functions}, 
mimicking collinear emission at leading power in transverse momenta. 
As an example, for final-state collinear radiation from a parent quark 
one defines
\beq
  \overline{u}_s (p) \, {\cal J}_{q, \, m}^{\{ \lambda \}} \big(\{ k \}; p, n \big)
  \, \equiv \, \bra{p, s; k_1, \lambda_1; \ldots; k_m, \lambda_m}  \overline{\psi} (0) \, 
  \Phi_{n} (0, \infty) \ket{0} \, ,
\label{qradjet}
\eeq
which in turn can be `squared' to yield the {\it collinear transition probability}
\beq
\label{qradjetsq}
  J_{q, \, m} \big( \{ k \}; l, p, n \big)
  & \equiv &
  \int d^d x \, {\rm e}^{{\rm i} l \cdot x} \sum_{\{ \lambda \}} \bra{0} 
  \overline{T} \Big[ \Phi_n (\infty, x) \, \psi(x) \Big] \ket{p, s; \{ k \}, \{ \lambda \}} \\ 
  && \hspace{4cm} \times \, 
  \bra{p, s; \{ k \}, \{ \lambda \}} T \Big[ \overline{\psi} (0) \, 
  \Phi_n (0, \infty) \Big] \ket{0} \, , \nonumber
\eeq
where the Furier transform has been introduced to fix the total final state 
momentum to $l$. Using completeness in this case yields the discontinuity
of a two-point function,
\beq
  \sum_{m = 0}^\infty \int d \Phi_{m+1} \, 
  J_{q, \, m} \big(\{ k \}; l, p, n \big) = 
  {\rm Disc} \bigg[ \! \int \! d^d x \, {\rm e}^{{\rm i} l \cdot x}
  \bra{0} T \Big[ \Phi_n (\infty, x) \psi(x) \overline{\psi} (0) 
  \Phi_n (0, \infty) \Big] \ket{0}
  \!  \bigg] .
\label{compcoll}
\eeq
which is also finite by power-counting. Note however that the collinear limit,
$l^2 \to 0$, is not intrinsic to the definition in \eq{qradjet} and must be taken
at a suitable stage in the calculation; furthermore, the collinear phase space
does not simply factorise as in the soft limit: in order to effect the cancellation
within the framework of subtraction, one needs to express the total (in general
off-shell) momentum $l^\mu$ in terms of an on-shell parent momentum 
$\bar{l}^\mu$, to be associated with the Born process. This is how phase-space 
mappings make their appearance within the factorisation framework.

\section{Strong ordering and refactorisation}
\label{Strong}

Once a soft or collinear kernel for multiple radiation has been computed in 
a uniform limit (with all particles becoming unresolved at the same rate), for
example by means of \eq{SoftFunc} or \eq{qradjetsq}, it is straightforward 
in practice to extract strongly-ordered limits, by performing a further Taylor 
expansion in the normal variables of the most unresolved particle. We believe 
however that it is useful to understand these limits theoretically, in terms of 
factorisation: this will contribute to the automatic construction of finite
combinations of real and mixed real-virtual corrections, by means of the
{\it completeness} technique highlighted in \secn{Facto}. The basic idea is to
treat eikonal form factors and jet functions as amplitudes in the presence of 
sources, and apply to them the same factorisation techniques that are used 
for the original scattering amplitudes. Here, as an example, we will focus on 
the soft sector: for related work in collinear limits, see~\cite{Braun-White:2022rtg,
Catani:2022sgr,Us}.

\subsection{Tree-level multiple soft emissions}
\label{Tree}

Consider first the emission of two soft gluons at tree level. The corresponding 
current was computed in~\cite{Catani:1999ss}, and the result is easily reproduced
by using \eq{softrad} at tree level. In the strongly-ordered limit, with soft momenta
$k_1^\mu \gg k_2^\mu$, the double soft-gluon current simplifies to
\beq
  \left[ J_{\CG}^{(0), \, \rm s.o.} \right]_{\mu_1 \mu_2}^{a_1 a_2} 
  \left( k_1, k_2; \beta_i \right) \, = \, 
  \left(J^{(0) \, a_2}_{\mu_2}(k_2) \, \delta^{a_1 a}+ {\rm i} g_s \, 
  f^{a_1 a_2 a} \, \frac{k_{1 , \, \mu_2}}{k_1\cdot k_2} \right)
  J^{(0)}_{\mu_1, \, a} (k_1) \, ,
\label{2sCGso}
\eeq
where the tree-level single soft-gluon current is given by the well-known 
expression
\beq 
  J_\mu^{(0) \, a} (k) \, = \, g_s \, \sum_{i = 1}^n \,
  \frac{\beta_{i,\, \mu}}{\beta_i \cdot k} \, T^a_i  \, .
\label{treesoftcurexp}
\eeq
The factorised structure of \eq{2sCGso} suggests that the two-gluon eikonal
form factor should also behave similarly at this order. Indeed, one finds the 
interesting factorisation
\beq
\label{softrads.o.}
  \Big[{\cal S}^{(0)}_{n; \, 1, \, 1} \Big]^{a_1 a_2}_{\{d_i e_i\}}
  \big(k_1, k_2; \beta_{i} \big) & \equiv &  
  \bra{k_2, a_2} \, 
  \Phi^{\,\, a_1 b}_{\beta_{k_1}} (0,\infty) 
  \prod_{i = 1}^{n} \Phi^{\quad \,\,\,\,\, c_i }_{\beta_{i}, \, d_i}(0,\infty) \, \ket{0} 
  \nn\\
  & & \hspace{1cm}
  \times \,
  \bra{k_1, b} \,  \prod_{i=1}^{n}
  \Phi_{\beta_i, \, c_i e_i}(0,\infty)  \ket{0} \Big|_{\rm tree} \nn \\
  & = & \left[ {\cal S}^{(0)}_{n+1,1} \right]^{a_2, \, a_1 b}_{\{d_i c_i\}}
  \big(k_2;  \beta_{k_1}, \beta_{i} \big) \, 
  \left[ {\cal S}^{(0)}_{n,1} \right]_{b, \, \{ c_i e_i \}} \! \big(k_1; \beta_{i} \big)\, ,
\eeq
where for clarity we displayed all colour indices, and where we adopted the 
notation ${\cal S}^{(0)}_{n; \, m_1, \ldots \, m_p}$ for the strongly-ordered 
emission of $p$ clusters of soft gluons, each containing $m_k$ gluons, 
$k = 1, \ldots, p$. Notice the non-trivial colour structure of \eq{softrads.o.}: 
the product on the {\it r.h.s.} is ordered, and the colour index of the $k_1$ 
Wilson line in 
the first factor is contracted with the colour index of the final-state gluon 
in the second factor. The physical interpretation is transparent: first, the 
original system of $n$ Wilson lines corresponding to the hard particles 
radiates the harder gluon with momentum $k_1$, corresponding to the 
second factor on the {\it r.h.s.} of \eq{softrads.o.}. This first gluon is much 
harder than the second one with momentum $k_2$, thus it turns into a 
Wilson line in the first factor (we could say that we now have a {\it Wilsonised}
gluon).  The augmented system of $\npo$ Wilson lines then radiates the 
softer gluon. One easily checks that \eq{softrads.o.} reproduces \eq{2sCGso},
contracted with appropriate polarisation vectors. Clearly, \eq{softrads.o.}
lends itself to a natural generalisation for the strongly-ordered emission
of any number of gluons, organised in clusters of generic multiplicity. 
As an example, consider the emission of three gluons with strongly-ordered
momenta $k_1^\mu \gg k_2^\mu \gg k_3^\mu$. \eq{softrads.o.} suggests 
the factorisation
\beq
  && \left[ {\cal S}_{n;1,1,1}^{(0)} \right]^{a_1a_2a_3}_{\{ f_i e_i \}}
  \left(k_1, k_2, k_3; \beta_{i} \right) \, \equiv \, 
  \left[ {\cal S}^{(0)}_{n+2,1} \right]^{a_3}_{\{ f_i d_i \}, \, a_1 b_1, \, a_2 b_2}
  \left[ {\cal S}^{(0)}_{n+1,1} \right]^{b_2}_{\{ d_i c_i \}, \, b_1 g_1}
  \left[ {\cal S}^{(0)}_{n,1} \right]^{g_1}_{\{ c_i e_i \}} \nn \\ 
  && \,\,\, = \, \,\langle k_3, a_3 | \,
  \Phi_{\beta_{k_1}}^{a_1 b_1}(0, \infty) \,
  \Phi_{\beta_{k_2}}^{a_2b_2} (0, \infty) 
  \prod_{i = 1}^n \Phi_{\beta_{i}}^{f_i d_i} (0, \infty) | 0 \rangle \, \,
  \langle k_2, b_2 | \,
  \Phi_{\beta_{k_1}}^{b_1g_1}(0, \infty) 
  \prod_{i=1}^n \Phi_{\beta_{i}}^{d_ic_i}(0, \infty) | 0 \rangle \nn \\
  && \hspace{4cm} \times \, \langle k_1,g_1 | \prod_{i=1}^n
  \Phi_{\beta_{i}}^{c_ie_i}(0, \infty) | 0 \rangle
  \Big|_{\rm tree} \, .
\label{softs.o.3}
\eeq
One can readily verify that \eq{softs.o.3} reproduces the 
strongly-ordered limit of the triple soft-gluon current, derived and 
discussed in~\cite{Catani:2019nqv,Colferai:2022jjf}: an explicit
calculation indeed gives the current
\beq
\label{3soft}
  \left[ J_{\rm CCT}^{(0), \, \rm s.o.} \right]_{\mu_1 \mu_2 \mu_3}^{a_1 a_2 a_3} 
  \left( k_1, k_2, k_3; \beta_i \right) & = &
  \Bigg[
  J^{\mu_3}_{a_3}(k_3) \, \delta^{a_1 b_1} \, \delta^{a_2 b_2}
  + \, {\rm i} g_s \, f^{a_1 a_3 b_1} \, \delta^{a_2 b_2} \,
  \frac{k_1^{\mu_3}}{k_1\cdot k_3} \\
  && \hspace{-2cm} + \, {\rm i} g_s \, f^{a_2 a_3 b_2} \, \delta^{a_1 b_1} \, 
  \frac{k_2^{\mu_3}}{k_2\cdot k_3}
  \Bigg] \, \Bigg[
  J^{\mu_2}_{b_2} (k_2) \, \delta^{b_1 c_1}
  + \, {\rm i} g_s \, f^{b_1 b_2 c_1} \, \frac{k_1^{\mu_2}}{k_1\cdot k_2}
  \Bigg] \, J_{c_1}^{\mu_1}(k_1) \, . \nn 
\eeq
It is not difficult to write down generalisations of Eqs.~(\ref{softrads.o.}) and 
(\ref{softs.o.3}) to more soft particles and more general clusterings, although
the resulting expressions are somewhat cumbersome: such expressions provide
direct and straightforward ways to compute the corresponding tree-level currents.
One should note also that the Wilson line matrix elements introduced above can
of course be evaluated to any order in perturbation theory, and one may ask
to what extent this tree-level factorisation generalises to higher orders. This question
in general remains open, but we will now examine how a one-loop factorisation of 
this kind impacts the construction of subtraction counterterms.

\subsection{One-loop refactorisation and subtraction counterterms}
\label{Loop}

Within the framework of the factorisation approach to subtraction~\cite{Magnea:2018ebr},
using the definition of local soft counterterms given by \eq{SoftFunc}, a natural 
candidate counterterm for real-virtual soft singularities is given by
\beq
  K_\npo^{\RV, \, {\rm s}} \, = \, {\cal A}^{(0) \, \dagger}_n \, S^{(1)}_{n, \, 1} 
  \, {\cal A}^{(0)}_n \, + \, \ldots  \, ,
\label{KRVtentS}
\eeq
where ${\cal A}_n^{(0)}$ is the Born amplitude, and $S^{(1)}_{n, \, 1}$ is the 
one-loop contribution to the single-radiative soft function, which is a colour 
matrix. The ellipsis contains terms that have no soft poles (but might still 
display soft phase-space singularities), or terms that are non-singular in
phase space, but might still have soft poles. In words, $S^{(1)}_{n, \, 1}$ 
accurately reproduces terms with joint soft poles and singular soft phase-space 
singularities of the complete squared matrix element. On the other hand, one 
sees from \eq{NNLOfin} that the soft poles of $K_\npo^{\RV,  \, {\rm s}}$ must 
cancel those arising from the integration of the strongly-ordered double-radiative 
soft counterterm $K^{\otwo, \, {\rm s}}_\npt$ over the softest-particle phase space. 
The refactorisation of the double-radiative tree-level soft function in the strongly-ordered 
limit, discussed in \secn{Tree} suggests an expression for the corresponding 
counterterm. One writes
\beq
  && K_\npt^{\otwo, \, {\rm s}} \, = \, {\cal A}^{(0) \, \dagger}_n \, S^{(0)}_{n, \, 1,1} 
  \, {\cal A}^{(0)}_n \nonumber \\ 
  && = \, {\cal A}^{(0) \, \dagger}_n \,
  \left[ {\cal S}_{n, \, 1}^{b, \, (0)} (\beta_i; k_1) \right]^\dagger \, 
  \left[ {\cal S}_{\npo, \, 1}^{\, a_2, \, a_1 b \, (0)} (\beta_i, \beta_{k_1}; k_2) \right]^\dagger \, 
  {\cal S}_{\npo, \, 1}^{\, a_2, \, a_1 c, (0)} (\beta_i, \beta_{k_1}; k_2) \,
  {\cal S}_{n, \, 1}^{c, \, (0)} (\beta_i; k_1) \, {\cal A}^{(0)}_n \nonumber \\
  && = \, {\cal A}^{(0) \, \dagger}_n \,
  \left[ {\cal S}_{n, \, 1}^{b, \, (0)} (\beta_i; k_1) \right]^\dagger \,
  S_{\npo, \, 1}^{\, b c, \, (0)} (\beta_i, \beta_{k_1}; k_2) \,
  {\cal S}_{n, \, 1}^{c, \, (0)} (\beta_i; k_1) \, {\cal A}^{(0)}_n \, . \nonumber
\label{Kotwotent}
\eeq
Note now that the dependence on the softest momentum $k_2$ is confined 
to the innermost factor in the product on the last line. One can then use the 
finiteness of \eq{complete} at one loop,
\beq
  S_{\npo, \, 0}^{\, b c, (1)} ( \beta_i, \beta_{k_1} ) \, + \,
  \int d \Phi_1 (k_2) \, S_{\npo, \, 1}^{\, b c, (0)} (\beta_i, \beta_{k_1}; k_2) 
  \, =  \, [{\rm finite \, in} \, d=4] \, ,
\label{complesoft}
\eeq
to propose what appears to be an alternative expression for the soft real virtual 
counterterm,
\beq
  K_\npo^{\RV, \, {\rm s}} \, = \, {\cal A}^{(0) \, \dagger}_n \, 
  \left[ {\cal S}_{n, \, 1}^{b, \, (0)} (\beta_i; k_1) \right]^\dagger \,
  S_{\npo, \, 0}^{\, b c, (1)} ( \beta_i, \beta_{k_1} ) \,
  {\cal S}_{n, \, 1}^{c, \, (0)} (\beta_i; k_1) \, {\cal A}^{(0)}_n
  \, + \, \ldots \, .
\label{KRV_Alt}
\eeq
The obvious questions are whether \eq{KRVtentS} and \eq{KRV_Alt} are
compatible, and whether they match a direct calculation of soft singularities
for the real-virtual squared matrix element. To answer these questions, we
turn once again to the factorisation properties of the soft function, this time
at the one-loop level. Treating the single-radiative eikonal form factor as
a scattering amplitude in the presence of Wilson-line sources suggests that 
it can be factorised in the form
\beq
  {\cal S}_{n, 1} (k; \beta_i) \, = \, \frac{{\cal J}_g (k, n)}{{\cal J}_{E, g} (\beta_k, n)} 
  \, \, {\cal S}_{\npo, 0} (\beta_k, \beta_i) \,\,  {\cal S}_{n, 1}^{\rm fin} (k; \beta_i) \, ,
\label{ampfac}
\eeq
which is reminiscent of the general form of infrared factorisation for virtual 
corrections to scattering amplitudes~\cite{Gardi:2009qi,Becher:2009qa,
Feige:2014wja,Agarwal:2021ais}: ${\cal J}_g (k, n)$ is a gluon jet function 
responsible for collinear divergences associated with the radiated gluon, 
${\cal J}_{E, g} (\beta_k, n)$ is its eikonal counterpart, responsible for the 
subtraction of soft-collinear poles, ${\cal S}_{\npo, 0}$ is the virtual soft function 
for the full set of $(\npo)$ particles, and ${\cal S}_{n, 1}^{\rm fin} (k; \beta_i)$ 
is IR finite. Expanding \eq{ampfac} to one loop, one finds
\beq
  {\cal S}_{n, 1}^{(1)} (k; \beta_i) \, = \, 
  {\cal S}_{\npo, 0}^{(1)} (\beta_k, \beta_i) \, {\cal S}_{n, 1}^{(0)} (k; \beta_i) \, + \,
  \left( {\cal J}_g^{(1)} (k, n) - {\cal J}_{E, g}^{(1)} (\beta_k, n) \right) \, 
  {\cal S}_{n, 1}^{(0)} (k; \beta_i) \, .
\label{softrefac1}
\eeq
Squaring \eq{softrefac1}, and retaining one-loop contributions, explains the 
relationship between the two candidate definitions of $K_\npo^{\RV, \, {\rm s}}$,
Eqs.~(\ref{KRVtentS}) and (\ref{KRV_Alt}): they differ by purely {\it hard collinear} 
contributions, arising from the soft-subtracted gluon jet. The two definitions are 
therefore consistent in the soft limit, as claimed. A non-trivial calculation of the 
soft limit of the real-virtual squared matrix element, and of the one-loop contribution 
to the single-radiative soft function, fully confirms these rather formal arguments. 
One finds~\cite{Us}
\beq
  S_{n,1}^{(1)}(k; \beta_i) \, = \, \bS{k} \, RV_\npo \, - \, 
  \frac{\as^2 \mu^{2 \epsilon}}{S_\epsilon} \, 
  \sum_{i > j}^n \frac{\beta_i \cdot \beta_j}{\beta _i \cdot k \, 
  \beta_j \cdot k} \, {\bf T}_i \cdot {\bf T}_j
  \left[ \, \sum_{m = 1}^n\frac{\gamma_m^{(1)}}{\eps} + \frac{b_0}{2 \eps}
  \right] \, ,
\label{SkRV}
\eeq
where $\gamma_m^{(1)}$ is the one-loop contribution to the collinear 
anomalous dimension, responsible for hard collinear poles, $S_\eps = 
(4 \pi {\rm e}^{- \gamma_E})^\eps$ is the standard $\overline{MS}$ factor,
and the calculation was performed for the bare soft function. \eq{SkRV}
states that the one-loop contribution to the radiative soft function fully 
captures the soft singularities of the real-virtual squared matrix element,
up to corrections that are proportional to purely collinear poles.

\section{Outlook}
\label{Out}

Strongly-ordered infrared limits are important ingredients for subtraction
algorithms, whose relevance and intricacy grow steeply at high orders,
and they have an interesting structure from the point of view of factorisation.
We have proposed tree-level factorisation formulas for strongly-ordered
soft limits, which naturally generalise to an arbitrary number of soft emissions, 
and match existing results. At loop level, we have presented evidence that a 
refactorisation of soft functions provides insights in the structure of real-virtual 
soft subtraction counterterms: indeed, applying finiteness constraints which 
follow from completeness sums like \eq{complete} successfully links strongly-ordered 
double radiation to real-virtual corrections. With appropriate adjustments, our 
results generalise to collinear limits, where multiple radiative jet functions 
refactorise into products of lower-order ones in strongly-ordered collinear limits. 
In the collinear case, however, a detailed implementation must tackle the issue 
of phase-space mappings. Further details will be presented in a forthcoming 
publication~\cite{Us}.


\acknowledgments

This research was partially supported by the Deutsche Forschungsgemeinschaft 
(DFG, German Research Foundation) under grant 396021762 - TRR 257, and
by the Italian Ministry of University and Research (MIUR), grant PRIN 20172LNEEZ. 
The work  of PT has received support from Compagnia di San Paolo, grant 
n. TORP\_S1921\_EX-POST\_21\_01.



\begin{thebibliography}{99}

\vspace{-1mm}
\bibitem{Frixione:1995ms}
S.~Frixione, Z.~Kunszt, and A.~Signer,
  {\em Nucl. Phys.} {\bf B467} (1996) 399--442,
  [\href{http://xxx.lanl.gov/abs/hep-ph/9512328}{{\tt hep-ph/9512328}}].

\vspace{-1mm}
\bibitem{Catani:1996vz}
S.~Catani and M.~H. Seymour,
  {\em Nucl. Phys.} {\bf B485} (1997) 291--419,
  [\href{http://xxx.lanl.gov/abs/hep-ph/9605323}{{\tt hep-ph/9605323}}].
  [Erratum: Nucl. Phys.B510,503(1998)].

\vspace{-1mm}
\bibitem{Nagy:2003qn} 
  Z.~Nagy and D.~E.~Soper, 
  {\em JHEP} {\bf 0309}, 055 (2003)
  [\href{http://xxx.lanl.gov/abs/hep-ph/0308127}{{\tt hep-ph/0308127}}].

\bibitem{Bevilacqua:2013iha}
G.~Bevilacqua, M.~Czakon, M.~Kubocz and M.~Worek,
JHEP \textbf{10} (2013), 204
[arXiv:1308.5605 [hep-ph]].

\vspace{-1mm}
\bibitem{GehrmannDeRidder:2005cm} 
  A.~Gehrmann-De Ridder, T.~Gehrmann and E.~W.~N.~Glover,
  {\em JHEP} {\bf 0509}, 056 (2005),
  [\href{http://xxx.lanl.gov/abs/hep-ph/0505111}{{\tt hep-ph/0505111}}].

\bibitem{Somogyi:2005xz} 
  G.~Somogyi, Z.~Trocsanyi and V.~Del Duca,
  {\em JHEP} {\bf 0506}, 024 (2005),
  [\href{http://xxx.lanl.gov/abs/hep-ph/0502226}{{\tt hep-ph/0502226}}].

\bibitem{Czakon:2010td}
  M.~Czakon,
  {\em Phys. Lett.} {\bf B693} (2010) 259--268,
  [\href{http://xxx.lanl.gov/abs/1005.0274}{{\tt 1005.0274}}].

\bibitem{Binoth:2000ps} 
  T.~Binoth and G.~Heinrich,
  {\em Nucl.\ Phys.\ B} {\bf 585}, 741 (2000),
  [\href{http://xxx.lanl.gov/abs/hep-ph/0004013}{{\tt hep-ph/0004013}}].

\bibitem{Anastasiou:2003gr} 
  C.~Anastasiou, K.~Melnikov and F.~Petriello,
  {\em Phys.\ Rev.\ D} {\bf 69}, 076010 (2004),
  [\href{http://xxx.lanl.gov/abs/hep-ph/0311311}{{\tt hep-ph/0311311}}].

\bibitem{Caola:2017dug}
  F.~Caola, K.~Melnikov, and R.~R{\"o}ntsch,
  {\em Eur. Phys. J.} {\bf C77}  (2017), no.~4 248,
  [\href{http://xxx.lanl.gov/abs/1702.01352}{{\tt 1702.01352}}].

\bibitem{Catani:2007vq}
  S.~Catani and M.~Grazzini,
  {\em Phys. Rev. Lett.} {\bf 98} (2007) 222002,
  [\href{http://xxx.lanl.gov/abs/hep-ph/0703012}{{\tt hep-ph/0703012}}].

\bibitem{Boughezal:2015dva} 
  R.~Boughezal, C.~Focke, X.~Liu and F.~Petriello,
  {\em Phys.\ Rev.\ Lett.} {\bf 115}, no. 6, 062002 (2015),
  [\href{http://xxx.lanl.gov/abs/1504.02131}{{\tt 1504.02131}}].

\bibitem{Cacciari:2015jma}
  M.~Cacciari, F.~A. Dreyer, A.~Karlberg, G.~P. Salam, and G.~Zanderighi,
  {\em Phys. Rev. Lett.} {\bf 115} (2015),
  no.~8 082002, [\href{http://xxx.lanl.gov/abs/1506.02660}{{\tt 1506.02660}}].

\bibitem{Sborlini:2016hat}
G.~F.~R. Sborlini, F.~Driencourt-Mangin, and G.~Rodrigo,
  {\em JHEP} {\bf 10} (2016) 162,
  [\href{http://xxx.lanl.gov/abs/1608.01584}{{\tt 1608.01584}}].

\bibitem{Herzog:2018ily}
  F.~Herzog,
  {\em JHEP} {\bf 1808}, 006 (2018),
  \href{http://xxx.lanl.gov/abs/1804.07949}{{\tt 1804.07949}}.

\bibitem{Magnea:2018hab} 
  L.~Magnea, E.~Maina, G.~Pelliccioli, C.~Signorile-Signorile, P.~Torrielli 
  and S.~Uccirati,
  {\em JHEP} {\bf 1812}, 107 (2018), Erratum: [JHEP {\bf 1906}, 013 (2019)]
  \href{http://xxx.lanl.gov/abs/1806.09570}{{\tt 1806.09570}}.

\bibitem{Capatti:2019ypt}
Z.~Capatti, V.~Hirschi, D.~Kermanschah and B.~Ruijl,
Phys. Rev. Lett. \textbf{123} (2019) no.15, 151602
[arXiv:1906.06138 [hep-ph]].

\bibitem{TorresBobadilla:2020ekr}
W.~J.~Torres Bobadilla, G.~F.~R.~Sborlini, P.~Banerjee, S.~Catani, 
A.~L.~Cherchiglia, L.~Cieri, P.~K.~Dhani, F.~Driencourt-Mangin, 
T.~Engel and G.~Ferrera, \textit{et al.}
Eur. Phys. J. C \textbf{81} (2021) no.3, 250
doi:10.1140/epjc/s10052-021-08996-y
[arXiv:2012.02567 [hep-ph]].

\bibitem{Agarwal:2021ais}
N.~Agarwal, L.~Magnea, C.~Signorile-Signorile and A.~Tripathi,
[arXiv:2112.07099 [hep-ph]].

\bibitem{Magnea:2018ebr} 
  L.~Magnea, E.~Maina, G.~Pelliccioli, C.~Signorile-Signorile, P.~Torrielli and 
  S.~Uccirati,
  {\em JHEP} {\bf 1812}, 062 (2018)
  \href{http://xxx.lanl.gov/abs/1809.05444}{{\tt 1809.05444}}.

\bibitem{Feige:2014wja}
I.~Feige and M.~D.~Schwartz,
Phys. Rev. D \textbf{90} (2014) no.10, 105020
doi:10.1103/PhysRevD.90.105020
[arXiv:1403.6472 [hep-ph]].

\bibitem{Magnea:2020trj}
L.~Magnea, G.~Pelliccioli, C.~Signorile-Signorile, P.~Torrielli and S.~Uccirati,
JHEP \textbf{02} (2021), 037
[arXiv:2010.14493 [hep-ph]].

\bibitem{UcciratiLL}
S.~Uccirati, contribution to these Proceedings.

\bibitem{DelDuca:2019ctm}
V.~Del Duca, N.~Deutschmann and S.~Lionetti,
JHEP \textbf{12} (2019), 129
[arXiv:1910.01024 [hep-ph]].

\bibitem{Braun-White:2022rtg}
O.~Braun-White and N.~Glover,
[arXiv:2204.10755 [hep-ph]].

\bibitem{Catani:2022sgr}
S.~Catani and P.~K.~Dhani,
[arXiv:2208.05840 [hep-ph]].

\bibitem{Us}
L.~Magnea, C.~Milloy, C.~Signorile-Signorile, P.~Torrielli, S.~Uccirati, 
in preparation.

\bibitem{Catani:1999ss} 
S.~Catani and M.~Grazzini,
{\em Nucl.\ Phys.\ B} {\bf 570}, 287 (2000),
[\href{http://xxx.lanl.gov/abs/hep-ph/9908523}{{\tt hep-ph/9908523}}].

\bibitem{Catani:2019nqv}    
S.~Catani, D.~Colferai and A.~Torrini,
JHEP \textbf{01} (2020), 118
[arXiv:1908.01616 [hep-ph]].

\bibitem{Colferai:2022jjf}
D.~Colferai, S.~Catani and A.~Torrini,  
[arXiv:2207.01717 [hep-ph]].

\bibitem{Gardi:2009qi}
E.~Gardi and L.~Magnea,
JHEP \textbf{03} (2009), 079
[arXiv:0901.1091 [hep-ph]].

\bibitem{Becher:2009qa}
T.~Becher and M.~Neubert,
JHEP \textbf{06} (2009), 081
[erratum: JHEP \textbf{11} (2013), 024]
[arXiv:0903.1126 [hep-ph]].
\end{thebibliography}
\end{document}